\documentclass[submission, copyright, creativecommons]{eptcs}
\providecommand{\event}{1st workshop on Architectures, Languages and Paradigms\\
	for IoT (ALP4IoT'17), September 18th, 2017, Turin, Italy.}

\usepackage{amsmath}
\usepackage{amssymb}
\usepackage{amsthm}
\usepackage[mathcal]{euscript}
\usepackage{stmaryrd}
\usepackage{xspace}
\usepackage{upquote}
\usepackage{listings}
\usepackage{color}
\usepackage{mathrsfs}
\usepackage{microtype}
\usepackage{url}
\usepackage{graphicx}
\usepackage{wrapfig}
\usepackage[strings]{underscore}

\newcommand{\mtch}{\ensuremath{\mathit{match}}}

\newcommand{\emptyseq}{\epsilon}

\newcommand{\trans}[1]{\stackrel{{#1}}{\rightarrow}}
\newcommand{\isEmpty}{\mathit{\epsilon}}

\newcommand{\ping}{\mathit{ping}}
\newcommand{\pong}{\mathit{pong}}

\newcommand{\Rule}[4]{\scriptstyle{\textrm{({#1})}}{\displaystyle\frac{#2}{#3}}\ #4}

\newcommand{\extrans}[1]{\stackrel{{#1}}{\rightarrowtail}}

\newcommand{\eventSet}{\mathcal{E}}
\newcommand{\eventTy}{\vartheta}

\newcommand{\ev}{e}
\newcommand{\evtr}{\bar{\ev}}

\newcommand{\sem}[1]{\llbracket{#1}\rrbracket}

\newcommand{\prefixop}{\mathbin{:}}
\newcommand{\orop}{\mathbin{\vee}}
\newcommand{\shuffleop}{\mathbin{|}}
\newcommand{\catop}{\mathbin{\cdot}}

\newcommand{\andop}{\mathbin{\wedge}}

\newcommand{\restrict}[2]{{#1}_{\setminus{#2}}}

\newcommand{\op}{\ensuremath{\mathit{op}}}

\newcommand{\subs}{\ensuremath{\sigma}}

\newcommand{\subsMerge}{\ensuremath{\cup}}

\newcommand{\dom}{\ensuremath{\mathit{dom}}}

\newcommand{\vals}{\ensuremath{\mathcal{V}}}

\newcommand{\avar}[1]{\mathsf{#1}}
\newcommand{\var}[2]{{<}\avar{#1}\mathbin{;}{#2}{>}}

\newcommand{\opent}{\ensuremath{\mathit{open}}}
\newcommand{\closet}{\ensuremath{\mathit{close}}}
\newcommand{\writet}{\ensuremath{\mathit{write}}}

\newcommand{\val}{\mathsf{val}}
\newcommand{\fd}{\mathsf{fd}}
\newcommand{\id}{\mathsf{id}}
\newcommand{\xv}{\mathsf{x}}

\newcommand{\funpre}{\mathit{fun\_pre}}
\newcommand{\funpost}{\mathit{fun\_post}}
\newcommand{\cbpre}{\mathit{cb\_pre}}
\newcommand{\cbpost}{\mathit{cb\_post}}
\newcommand{\cb}{\mathit{cb}}

\definecolor{mygreen}{rgb}{0,0.6,0}
\definecolor{mygray}{rgb}{0.5,0.5,0.5}
\definecolor{mymauve}{rgb}{0.58,0,0.82}
 
\definecolor{darkgray}{rgb}{.4,.4,.4}
\definecolor{purple}{rgb}{0.65, 0.12, 0.82}
 
\lstdefinelanguage{JavaScript}{
keywords={typeof, new, true, false, catch, function, return, null, catch, switch, var, if, in, while, do, else, case, break,const},
keywordstyle=\color{blue}\bfseries,
ndkeywords={class, export, boolean, throw, implements, import, this},
ndkeywordstyle=\color{darkgray}\bfseries,
identifierstyle=\color{black},
sensitive=false,
comment=[l]{//},
morecomment=[s]{/*}{*/},
commentstyle=\color{purple}\ttfamily,
stringstyle=\color{mygreen}\ttfamily,
morestring=[b]',
morestring=[b]"
}
 
\newcommand*\inlinejs[1]{\lstinline[language=JavaScript]{#1}}

\title{Towards Runtime Monitoring of Node.js and Its Application to the Internet of Things}
\author{Davide Ancona, Luca Franceschini, Giorgio Delzanno, Maurizio Leotta \\ Marina Ribaudo, Filippo Ricca
\institute{DIBRIS, University of Genoa}
\email{\{\it name.surname\}@unige.it, luca.franceschini@dibris.unige.it}
}
\def\titlerunning{Towards Runtime Monitoring of Node.js and Its Application to the Internet of Things}
\def\authorrunning{Ancona, Franceschini, Delzanno, Leotta, Ribaudo \& Ricca}

\begin{document}
\maketitle
\begin{abstract}
In the last years Node.js has emerged as a framework particularly suitable for implementing lightweight IoT applications, thanks to its underlying asynchronous event-driven, non blocking {I/O} model.
However, verifying the correctness of programs with asynchronous nested callbacks is
quite difficult, and, hence, runtime monitoring can be a valuable support to tackle such a complex task. 

Runtime monitoring is a useful software verification technique that complements static analysis and testing, but has not been yet fully explored in the context of Internet of Things (IoT) systems. Trace expressions have been 
successfully employed for runtime monitoring in widespread multiagent system platforms.
Recently, their expressive power has been extended to allow parametric specifications on data that can be captured and monitored only at runtime.
Furthermore, they can be language and system agnostic, through the notion of event domain and type. 
This paper investigates the use of parametric trace expressions as a first step towards runtime monitoring of programs developed in Node.js and Node-RED, a flow-based IoT programming tool built on top of Node.js.
Runtime verification of such systems is a task that mostly seems to have been overlooked so far in the literature.


A prototype implementing the proposed system for Node.js, in order  to dynamically check with trace expressions the correct usage of API functions, is presented. 
The tool exploits the dynamic analysis framework Jalangi for monitoring Node.js 
programs and allows detection of errors that would be difficult to catch with other techniques.
Furthermore, it offers a simple REST interface which can be exploited for runtime verification of Node-RED components, and, more generally, IoT devices.
\end{abstract}

\section{Introduction}

Node.js\footnote{\url{https://nodejs.org}} is a platform for server-side JavaScript applications which is based on an event-driven non blocking I/O model, expressly designed to support easy and rapid development of lightweight servers able to manage a
considerable number of simultaneous requests.
For these features, and the huge module open ecosystem and
large associated community of developers, Node.js has emerged as a convenient platform for developing IoT applications.

Node-RED\footnote{\url{https://nodered.org/}} is a visual tool built on top of Node.js to easily wire together mobile devices, sensors, and, more in general, 
real-world events, and to add the necessary logic to integrate them with databases, messaging systems, and cloud platforms.
Thanks to its lightweight implementation based on Node.js, it can run on devices at the edge of the network, such as the Raspberry Pi\footnote{\url{https://www.raspberrypi.org/}}. 
Node-RED turns out to be extremely useful also for implementing mocks for sensors and mobile devices for 
acceptance testing of IoT systems \cite{ENWOT17}.
For all these reasons, it is an ideal solution for rapid and effective development of innovative solutions to 
the challenges of the IoT technical revolution.   

To fully exploit its characteristics, Node.js strongly relies on asynchronous and continuation passing style programming: I/O operations are
performed through calls to asynchronous functions where a callback must be passed to specify
how the computation continues once the called I/O operation completed asynchronously \cite{DelzannoNode}.  
Indeed, the Node.js execution model consists of a main \emph{event loop} which is run on a single-threaded process. Despite this fact,
with asynchronous programming very subtle bugs can occur as in concurrent programs. Furthermore, 
the dynamic nature of JavaScript makes error detection even harder. Finally, in IoT scenarios which typically rely on the 
interaction of several distributed components, ensuring the correctness of a system becomes a challenging task \cite{ENWOT17}.    

Runtime verification~\cite{Leucker2009293} is a lightweight verification technique complementing software testing and formal verification. 
Dynamic checking of the correct behavior of a system is performed by a \emph{monitor} which verifies that the trace of events captured at runtime
is compliant with the expected behavior expressed with a specification formalism, thus essentially solving a ``word problem'' 
(that is, deciding whether a word is included in a language). This approach has been successfully employed in the context of distributed applications such as Web applications~\cite{DBLP:conf/rv/HalleV10,5427377} and multi-agent systems~\cite{AnconaEtAl12}, and of real-time systems \cite{LARVA}. Runtime verification avoids the complexity of traditional formal verification approaches at the expenses of less coverage by analyzing only a finite set of execution traces, and works directly with the actual system, thus scaling up very well~\cite{AnconaEtAl16}.
As opposite to software testing, runtime verification offers the possibility to recover from detected violations with suitable 
handlers \cite{GlobalProtocolsAAMAS15}.

Trace expressions have been inspired by previous work on behavioral types \cite{AnconaEtAl12,Betty16} to specify and
verify at runtime the correctness of interaction protocols in multiagent systems.
They have been successfully adopted for several purposes and applications in the  
context of different multiagent system platforms \cite{IDC2014}.
Recently, they have been extended \cite{AnconaParametric2017}
to allow specifications to be \emph{parametric} \cite{RV-Monitor14}
in data that can be captured and monitored only at runtime. Thanks to this extension, specifications which, for instance, depend on the values exchanged by objects 
through methods, or on the dynamically evolving collection of objects or resources available at runtime,
can be suitably modeled, and the number of correct programming patterns that can be specified is significantly enlarged. 

This paper investigates the use of trace expressions as a first step towards runtime monitoring of programs developed in Node.js and Node-RED, a flow-based IoT programming tool built on top of Node.js. 
By inspecting the Node.js API reference documentation, we have identified several patterns that
require calls to asynchronous functions and, possibly, to their corresponding callbacks, to occur in a correct order to prevent the application to exhibit unsafe and unpredictable behavior. Such patterns can be succinctly defined by means of trace expressions.
The tool exploits the dynamic analysis framework Jalangi~\cite{Jalangi} for monitoring Node.js programs.
Furthermore, it offers a simple REST interface which can be exploited for runtime verification of Node-RED components.
To the best of our knowledge, no other similar approaches for runtime verification of Node.js systems have been proposed in the literature.

After providing the necessary background concerning trace expressions in Section~\ref{sec:trace-exp},
we turn to consider examples of Node.js code with subtle bugs and show how runtime monitoring through trace expressions 
can help to detect such bugs (Section~\ref{sec:node}). The implementation details of our prototype tool
are discussed in Section~\ref{sec:impl}, together with examples of its use for monitoring Node.js applications and Node-RED flows.
Finally, Section~\ref{sec:conclu} concludes and outlines directions for future work.  

\section{Parametric trace expressions}\label{sec:trace-exp}


A trace expression defines the set of all possible event traces that can be correctly observed
during the execution of the program under monitoring, thus serving as a specification.
Its semantics is defined by a labeled transition system
where labels correspond to the events of the system that have to be monitored: if a trace expression
$\tau$ rewrites with event $\ev$ into a new trace expression $\tau'$, this means that the occurrence of
$\ev$ corresponds to a correct behavior according to $\tau$, and $\tau'$ represents the new state to be monitored in the continuation of the system; otherwise, an anomalous behavior is detected.

The rewriting rules defining the semantics of trace expressions can be directly turned into an algorithm for event trace 
recognition.
In this way, a monitor for dynamically verifying the correct behavior of a system specified by a trace expression $\tau$  
can be simply obtained from $\tau$, and from the implementation of the labeled transition system.

To make trace expressions language and system agnostic, the notions of \emph{event domain} and \emph{type}
are introduced.
Abstractly, an event domain consists of a set of events $\eventSet$ together with their semantics and structure.
More operationally, an event domain corresponds to the infrastructure required for capturing and conveniently representing at runtime
the set of events that are emitted during the execution of the system, and that have to be monitored to ensure its
correct functioning.
Typically, an event domain is implemented through code instrumentation.

As we will see in the rest of the paper, a possible example of an event domain supporting Node.js consists of calls to asynchronous functions and the execution of their associated callbacks.
Such events have useful information associated with them, like the name of the functions that is being invoked (if not anonymous), its parameters and the returned value.
Typical events for an event domain supporting Node-RED correspond to messages sent with several protocols (HTTP, WebSocket, MQTT, etc.) and may contain information such as the payload, the timestamp, the topic, the sender, and the receiver of the message. 

Events are abstracted by event types which play two important roles:
\begin{itemize}
\item They provide a simple language that allows trace expressions to be defined 
independently of the underlying  event domains. In fact,
the trace expression language is stratified in two different layers.
The basic layer  models an event domain and consists of a simple language of terms corresponding to event types.
The top layer defines several operators, whose semantics is independent from the notion of event domain,
for building trace expressions on top of event types.
\item They enhance the expressive power of trace expressions, by allowing the definition of particular sets of events 
which are useful for monitoring purposes. The semantics of event types is defined by the generic function $\mtch$
which specifies how events match event types. The rewriting rules of the labeled transition system are parametric 
in the function $\mtch$ which is simply assumed to be given. No assumptions are needed on the language 
used for defining $\mtch$. In practice, we have found convenient to define
$\mtch$ as a Prolog predicate (see Section~\ref{sec:impl}).
\end{itemize}

An event type $\eventTy$ is a term which can contain free variables. The semantics of $\eventTy$
is determined by $\mtch$ as follows:
we say that $\mtch(\ev,\eventTy)$ \emph{succeeds} iff there exists a substitution
$\subs$ s.t. $\mtch(\ev,\eventTy)=\subs$, otherwise we say that  $\mtch(\ev,\eventTy)$ \emph{fails}.
$\subs$ is a finite partial map whose domain $\dom(\subs)$ must coincide with the set of variables in $\eventTy$.
Its codomain is a universe of values $\vals$ which depends on the considered event domain.

In the following, we provide some examples of event types associated with the event domain we have implemented for Node.js and
that is further specified in Sections~\ref{sec:node} and \ref{sec:impl}.
One type of event that is useful to trace in Node.js (but also in other programming languages) consists in calls to and 
returns from functions.
Correspondingly, an event domain may include the following two basic event types:
\begin{itemize}
\item $\funpre(\mathit{name,args})$ matches all events corresponding to a call to a function 
with name $\mathit{name}$, and list of arguments $\mathit{args}$. For instance, the event type
\lstinline[mathescape=true,basicstyle=\ttfamily\normalsize]{$\funpre($'fs.writeSync'$,$[9,'hello']$)$} matches all calls to the function \lstinline[basicstyle=\ttfamily\normalsize]{fs.writeSync}, with arguments
\lstinline[basicstyle=\ttfamily\normalsize]{9} and \lstinline[basicstyle=\ttfamily\normalsize]{'hello'}, respectively. Analogously, \lstinline[mathescape=true,basicstyle=\ttfamily\normalsize]{$\funpre($'fs.writeSync'$,\avar{args})$}
matches all calls to the function \lstinline[basicstyle=\ttfamily\normalsize]{fs.writeSync}, with any list of arguments. In this case, a successful match
will produce the appropriate substitution associating the actual list of arguments with the variable $\avar{args}$.
\item $\funpost(\mathit{name,args,ret})$ matches all events corresponding to a return from a function 
with name $\mathit{name}$, list of arguments $\mathit{args}$, and returned value $\mathit{ret}$. For instance, the event type
\lstinline[mathescape=true,basicstyle=\ttfamily\normalsize]{$\funpost($'fs.openSync'$,$['tmp.txt','w']$,$9$)$} matches all returns from the function \lstinline[basicstyle=\ttfamily\normalsize]{fs.openSync}, 
with arguments \lstinline[basicstyle=\ttfamily\normalsize]{'tmp.txt'} and \lstinline[basicstyle=\ttfamily\normalsize]{'w'}, and returned value \lstinline[basicstyle=\ttfamily\normalsize]{9}, respectively.
\end{itemize}

A parametric trace expression $\tau$ 
is a regular term\footnote{Since regular terms can be expressed through a finite number of syntactic equations, no explicit recursion operator is needed.} built on top of the following operators\footnote{Infix binary operators associate from left, and are listed in decreasing order of precedence, that is, the first operator has the highest precedence.}: 
\begin{tabbing}
\qquad \= $\bullet$ $\emptyseq$ (empty trace)
\qquad \= $\bullet$ $\eventTy\prefixop\tau$ (\emph{prefix})  
\qquad \= $\bullet$ $\tau_1\catop\tau_2$ (\emph{concatenation}) 
\qquad \= $\bullet$ $\tau_1\andop \tau_2$ (\emph{intersection}) \\
\> $\bullet$ $\tau_1\orop \tau_2$ (\emph{union})
\> $\bullet$ $\tau_1\shuffleop \tau_2$ (\emph{shuffle}, a.k.a. \emph{interleaving})
\>\> $\bullet$ $\var{x}{\tau}$ (\emph{binder})
\end{tabbing}
 The trace expression $\subs\tau$ obtained from $\tau$ by substituting all free occurrences of $\xv\in\dom(\subs)$ in $\tau$ with $\subs(\xv)$, is coinductively defined 
as follows:
$$
\begin{array}{ccc}
\subs(\eventTy\prefixop\tau)=(\subs\eventTy)\prefixop(\subs\tau) &
\subs(\tau_1 \mathbin{\op} \tau_2)=(\subs \tau_1)\mathbin{\op}(\subs \tau_2) \mbox{ for }\op\in\{\orop,\andop,\shuffleop,\catop\} &
\subs(\var{x}{\tau})=\var{x}{\restrict{\subs}{\xv}\tau}
\end{array}
$$

The labeled transition system for parametric trace expressions can be found in Figure~\ref{param-trans-fig}.
\begin{figure*}[t]
\begin{center}
\begin{math}
\begin{array}{c}



\Rule{main}
{\tau\extrans{\ev}\tau';\emptyset}
{\tau\trans{\ev}\tau'}
{}
\qquad
\Rule{prefix}
{}
{\eventTy\prefixop\tau\extrans{\ev}\tau;\subs}
{
  \subs=\mtch(\ev,\eventTy)
}
\qquad
\Rule{or-l}
{\tau_1\extrans{\ev}\tau'_1;\subs}
{\tau_1\orop\tau_2\extrans{\ev}\tau'_1;\subs}
{}
\qquad
\Rule{or-r}
{\tau_2\extrans{\ev}\tau'_2;\subs}
{\tau_1\orop\tau_2\extrans{\ev}\tau'_2;\subs}
{}
\\[3ex]
\Rule{and}
{\tau_1\extrans{\ev}\tau'_1;\subs_1\quad\tau_2\extrans{\ev}\tau'_2;\subs_2}
{\tau_1\andop\tau_2\extrans{\ev}\tau'_1\andop\tau'_2;\subs}
{\subs=\subs_1\subsMerge\subs_2}
\qquad
\Rule{shuffle-l}
{\tau_1\extrans{\ev}\tau'_1;\subs}
{\tau_1\shuffleop\tau_2\extrans{\ev}\tau'_1\shuffleop\tau_2;\subs}
{}
\qquad
\Rule{shuffle-r}
{\tau_2\extrans{\ev}\tau'_2;\subs}
{\tau_1\shuffleop\tau_2\extrans{\ev}\tau_1\shuffleop\tau'_2;\subs}
{}
\\[3ex]
\Rule{cat-l}
{\tau_1\extrans{\ev}\tau'_1;\subs}
{\tau_1\catop\tau_2\extrans{\ev}\tau'_1\catop\tau_2;\subs}
{}
\qquad
\Rule{cat-r}
{\tau_2\extrans{\ev}\tau'_2;\subs}
{\tau_1\catop\tau_2\extrans{\ev}\tau'_2;\subs}
{\isEmpty(\tau_1)}
\qquad
\Rule{var-t}
{\tau\extrans{\ev}\tau';\subs}
{\var{\xv}{\tau}\extrans{\ev}\subs\tau';\restrict{\subs}{\xv}}
{\xv\in\dom(\subs)}
\\[3ex]
\Rule{var-f}
{\tau\extrans{\ev}\tau';\subs}
{\var{\xv}{\tau}\extrans{\ev}\var{\xv}{\tau'};\subs}
{\xv\not\in\dom(\subs)}
\qquad
\Rule{$\isEmpty$-empty}
{}
{\isEmpty(\emptyseq)}
{}
\qquad
\Rule{$\isEmpty$-var}
{\isEmpty(\tau)}
{\isEmpty(\var{\xv}{\tau})}
{}
\qquad
\Rule{$\isEmpty$-or-l}
{\isEmpty(\tau_1)}
{\isEmpty(\tau_1\orop\tau_2)}
{}
\\[3ex]
\Rule{$\isEmpty$-or-r}
{\isEmpty(\tau_2)}
{\isEmpty(\tau_1\orop\tau_2)}
{}
\qquad
\Rule{$\isEmpty$-others}
{\isEmpty(\tau_1)\quad\isEmpty(\tau_2)}
{\isEmpty(\tau_1 \mathbin{\op} \tau_2)}
{\op\in\{\shuffleop,\catop,\andop\}}


\end{array}
\end{math}




\end{center}
\caption{Transition system for parametric trace expressions}\label{param-trans-fig}
\end{figure*}

We denote with $\emptyset$ the substitution with the empty domain.
The equality $\subs=\subs_1\subsMerge\subs_2$ holds iff $\dom(\subs)=\dom(\subs_1)\cup\dom(\subs_2)$, and
for all $\xv\in\dom(\subs)$, $\subs(\xv)=\subs_1(\xv)$ if $\xv\in\dom(\subs_1)$, and $\subs(\xv)=\subs_2(\xv)$ if $\xv\in\dom(\subs_2)$ (hence,
$\subs_1$ and $\subs_2$  must coincide on $\dom(\subs_1)\cap\dom(\subs_2)$).
We denote with $\restrict{\subs}{\xv}$ the substitution where $\xv$ is removed from its domain: 
$\restrict{\subs}{\xv}=\subs'$ iff $\dom(\subs')=\dom(\subs)\setminus\{\xv\}$ and
for all $\xv\in\dom(\subs')$, $\subs'(\xv)=\subs(\xv)$. The notation $\subs\eventTy$ denotes the event type obtained from $\eventTy$ by substituting all occurrences of $\xv\in\dom(\subs)$ in $\eventTy$ with $\subs(\xv)$.

Rule (main) defines the transition relation $\tau\trans{\ev}\tau'$ in terms
of the auxiliary relation $\tau\extrans{\ev}\tau';\subs$, 
which returns the substitution $\subs$ computed during the transition
step. This rule can be applied only if the computed substitution is empty, since
trace expressions are not allowed to contain free variables.



In rule (prefix) a transition step is possible only when the current event $\ev$ matches the event type $\eventTy$, and the computed substitution is returned by the $\mtch$ function.

Rules for union and shuffle are straightforward: union corresponds to the choice between
$\tau_1$ and $\tau_2$, while shuffle consists of the interleaving of $\tau_1$ and $\tau_2$.

For concatenation the auxiliary judgment $\isEmpty(\tau)$ (defined in the same Figure) is needed
for identifying the trace expression that accepts the empty trace $\emptyseq$. More precisely,
rule (cat-r) states that $\tau_1\cdot\tau_2$ accepts a trace $\evtr$ if $\tau_1$
accepts the empty trace (side condition $\isEmpty(\tau_1)$), and $\tau_2$ accepts $\evtr$.

In rule (and) the side condition requires that the substitutions $\subs_1$ and $\subs_2$ computed for $\tau_1$ and $\tau_2$ must coincide on
the intersection of their domains.
The final substitution $\subs$ is obtained by merging $\subs_1$ and $\subs_2$. 
For instance, if \lstinline[mathescape=true,basicstyle=\ttfamily\normalsize]{$\tau=\funpost(\avar{name},\avar{args},$9$)\prefixop\tau_1
\andop\funpost(\avar{name},$['tmp.txt','w']$,\avar{ret})\prefixop\tau_2$}
and the event $\ev$ corresponds to the return from the call \lstinline[basicstyle=\ttfamily\normalsize]{fs.openSync('tmp.txt','w')} with
value \lstinline[basicstyle=\ttfamily\normalsize]{9},  then we have  
\begin{flushleft}
\qquad\lstinline[mathescape=true,basicstyle=\ttfamily\normalsize]{$\funpost(\avar{name},\avar{args},$9$)\prefixop\tau_1\extrans{\ev}\tau_1;\{\avar{name}\mapsto$'fs.openSync'$, \avar{args}\mapsto$['tmp.txt','w']$\}$}\\
\qquad\lstinline[mathescape=true,basicstyle=\ttfamily\normalsize]{$\funpost(\avar{name},$['tmp.txt','w']$,\avar{ret})\prefixop\tau_2\extrans{\ev}\tau_2;\{\avar{name}\mapsto$'fs.openSync'$, \avar{ret}\mapsto$9$\}$}
\end{flushleft}
therefore \lstinline[mathescape=true,basicstyle=\ttfamily\normalsize]{$\tau\extrans{\ev}\tau_1\andop\tau_2;\{\avar{name}\mapsto$'fs.openSync'$, \avar{args}\mapsto$['tmp.txt','w']$,\avar{ret}\mapsto$9$\}$}.

Two rules are required for the $\var{x}{\tau}$ construct. Rule (var-t) is needed 
when variable $\xv$ is contained in the domain of the computed substitution $\subs$. 
$\subs$ is applied to the trace expression $\tau'$ in which $\tau$ rewrites to, the binder 
is removed and $\xv$ is removed from the domain of the computed substitution ($\restrict{\subs}{\xv}$). 
Rule (var-f) is required when variable $\xv$ is not contained in the domain of the computed substitution $\subs$: 
the binder is not removed, and the computed substitution coincides with $\subs$. 

Rules for the auxiliary predicate $\isEmpty(\_)$ are straightforward.

The semantics $\sem{\tau}$ of a trace expression $\tau$ is defined in terms of the transition relation 
$\trans{\ev}$, and the predicate $\isEmpty(\_)$:

\begin{itemize}
\item The empty trace $\emptyseq$ belongs to $\sem{\tau}$ iff  $\isEmpty(\tau)$ is derivable.
\item If $\evtr$ is a non empty, possibly infinite event trace, then $\evtr$ belongs to
$\sem{\tau}$ iff there exists a possibly infinite reduction sequence starting from $\tau$ for $\evtr$, that is, $\tau=\tau_0\trans{\ev_1}\tau_1\ldots\tau_{n-1}\trans{\ev_n}\tau_n\ldots$, and $\evtr=\ev_1\ldots\ev_n\ldots$.  
\end{itemize}






\section{Runtime monitoring of Node.js applications}\label{sec:node}

In this section we will provide simple examples of Node.js code to show how
asynchronous programming can introduce subtle bugs in Node.js applications, and how some of these bugs can be detected
with the help of runtime monitoring. To this aim, we will define trace expressions to ensure the safe use
of the asynchronous functions provided by the \lstinline{fs} module. Such trace expressions are based
on an event domain able to deal with calls to and returns from asynchronous functions, and their associated callbacks.

We start by showing a simple Node.js example which manipulates files through the \lstinline{fs} module in a synchronous way:
\begin{lstlisting}[belowskip=-0.8 \baselineskip]
const fs=require('fs')
var fd=fs.openSync('tmp.txt','w')
fs.writeSync(fd,'Hello world!\n')
fs.closeSync(fd)
\end{lstlisting} 

While for Node.js applications the use of asynchronous functions is more customary, for simplicity, 
in this first example we use the synchronous version of the
open, write and close
functions for manipulating files. In this case, the expected correct pattern for writing data on a file is the standard one: 
open the file, write data on it, possibly several times, then close it. This behavior can be defined by the following trace expression $T$:
$$
\begin{array}{l}
T=\emptyseq\orop\opent\prefixop W \qquad
W=\writet\prefixop W \orop \closet\prefixop\emptyseq
\end{array}
$$

For simplicity, we have abstracted away several details, and the event types in the trace expression 
refer to the corresponding operations performed on a single file, hence, there is a unique
implicit file descriptor, and the specification is
not parametric. To make it parametric, a binder needs to be used to introduce the variable $\fd$
which is instantiated at runtime with the file descriptor returned by the 
open operation:
$$
\begin{array}{l}
PT=\emptyseq\orop\var{\fd}{\opent(\fd)\prefixop(PW \shuffleop PT)}\qquad
PW=\writet(\fd)\prefixop PW \orop \closet(\fd)\prefixop\emptyseq
\end{array}
$$

The event type $\opent(\fd)$ captures all events corresponding to a return from \lstinline{fs.openSync} with value $\fd$. In other words, by using the event types introduced in the
previous section, the set of events matched by $\opent(\fd)$
coincides with the set of events matched by \lstinline[mathescape=true,basicstyle=\ttfamily\normalsize]{$\funpost($'fs.openSync'$,[\mathit{fname}$,'w'$],\fd)$}
for all strings $\mathit{fname}$. The other 
event types $\writet(\fd)$ and $\closet(\fd)$ are defined analogously.

The main trace expression $PT$ is recursively defined and the shuffle operator is needed to correctly taking into account the possibility
that several files can be opened and manipulated simultaneously. Shuffle expresses the fact that operations on
different file descriptors can be safely interleaved.

To give an example of reduction, let us assume that the following two subsequent events are captured (or, in other words, the following events are received by the monitor):
\lstinline[mathescape=true,basicstyle=\ttfamily\normalsize]{fs.openSync($\mathit{fname_1}$,'w')} returns file descriptor 9,
\lstinline[mathescape=true,basicstyle=\ttfamily\normalsize]{fs.openSync($\mathit{fname_2}$,'w')} returns file descriptor 10. Then, the trace expression $PT$ defined above
successfully reduces with two transition steps into the following trace expression $PT'$:
$$
\begin{array}{l}
PT'=PW_1\shuffleop PW_2\shuffleop PT\quad
PW_1=\writet(9)\prefixop PW_1 \orop \closet(9)\prefixop\emptyseq \quad
PW_2=\writet(10)\prefixop PW_2 \orop \closet(10)\prefixop\emptyseq 
\end{array}
$$
In other words, the program is correct with respect to the given specification.

The previous code example does not exploit the advantages offered by Node.js in terms of performances, since synchronous operations are used.
Let us now consider a more complex asynchronous version:
\begin{lstlisting}[belowskip=-0.8 \baselineskip]
const fs=require('fs')
fs.open('tmp.txt','w',(err,fd)=>{
    if(!err)
        fs.write(fd,'Hello world!\n',()=>fs.close(fd,()=>{}))
})
\end{lstlisting} 

In comparison to the synchronous version, this code is definitely less readable, and chances to 
introduce bugs are higher.
In this case the code is correct, and exhibits three callbacks
defined by arrow functions (lambda expressions in the JavaScript jargon) at three different
nesting levels: the callback passed to \lstinline[mathescape=true,basicstyle=\ttfamily\normalsize]{open} which is called as soon as 
the file has been opened, with an error object\footnote{For sake of simplicity, we have removed all code that would be necessary for correctly handling errors.} and the file descriptor associated with the newly opened file passed
as arguments. The callback passed to \lstinline[mathescape=true,basicstyle=\ttfamily\normalsize]{write} which, for simplicity, ignores its arguments and calls
\lstinline[mathescape=true,basicstyle=\ttfamily\normalsize]{close}, and, at the deepest level, the callback passed to \lstinline[mathescape=true,basicstyle=\ttfamily\normalsize]{close}, which does not perform any operation.

If we observe the events matching the $\funpost$ event type while running the code fragment above, we get the following sequence:
return from \lstinline[mathescape=true,basicstyle=\ttfamily\normalsize]{fs.open}, return from its associated callback, return from \lstinline[mathescape=true,basicstyle=\ttfamily\normalsize]{fs.write},
return from its associated callback, return from \lstinline[mathescape=true,basicstyle=\ttfamily\normalsize]{fs.close}, and, finally, return from its associated callback.
This is, indeed, the correct sequence of expected events: a callback associated with an asynchronous operation on
a file descriptor  needs to be executed \textbf{before} a subsequent asynchronous operation is performed on the same
file descriptor. For instance, if a monitor detects a return from 
\lstinline[mathescape=true,basicstyle=\ttfamily\normalsize]{fs.close($\mathit{fd}$)} before a return from the callback of 
\lstinline[mathescape=true,basicstyle=\ttfamily\normalsize]{fs.write($\mathit{fd}$,$\mathit{data}$)}, then it has to signal an anomaly, since
the monitored program is trying to close a file before a write operation on it is completed.

A common source of bugs in Node.js consists in using asynchronous functions as they were synchronous. 
This may be reasonable in situations where a controlled number of operations can be safely interleaved, as happens, for instance,
if a program has to send a bunch of HTTP requests to different servers, or even to the same server, providing that
they can be safely handled in any order. But there are many other situations where extreme care is needed
when calling asynchronous functions to avoid subtle bugs. Unfortunately, this is not always explicitly stated in
the Node.js API reference documentation. There are, however, cases where the documentation is clear about this issue:
for instance, one can find the following notice regarding the specification of \lstinline[mathescape=true,basicstyle=\ttfamily\normalsize]{fs.write}:
\emph{``Note that it is unsafe to use \lstinline[mathescape=true,basicstyle=\ttfamily\normalsize]{fs.write} multiple times on the same file without waiting for the callback''.}

Let us consider, for instance, the following unsafe example\footnote{The \texttt{**} notation stands for exponentiation.}:
\begin{lstlisting}[belowskip=-0.8 \baselineskip]
const fs=require('fs')
const limit=2**24
fs.open('tmp.txt','w',(err,fd)=>{
    if(!err)
        for(let i=0;i<limit;i++)
            fs.write(fd,i+'\n',()=>{})
    fs.close(fd,()=>{})
})
\end{lstlisting}  

The behavior of such a program is non deterministic, and the expected correct order of the integers written on the file is not ensured.
Even when the programmer is deliberately fine with this non deterministic behavior, this program will exhibit
serious heap memory problems if the constant \lstinline[mathescape=true,basicstyle=\ttfamily\normalsize]{limit} holds values which are large enough:
either the program crashes\footnote{We have experienced this behavior for \texttt{limit}=$2^{24}$ 
by running the code above with Node.js v7.2.1 on an Intel(R) Core(TM) i7-4510U CPU at 2.00GHz with a 16GB RAM, getting the error after several minutes.} after some time with a fatal error due to JavaScript heap out of memory, or successfully terminates, but it exhibits very poor performance.

To guarantee that functions are called only after some callbacks are completed, we have to refine the event domain presented
in the previous section to be able to track
callbacks associated with function calls. To this aim, coupling of calls to asynchronous functions and
their corresponding callbacks is achieved by emitting a unique identifier associated with both. Accordingly, the event types 
$\funpre$ and $\funpost$ are extended: $\funpre(\mathit{name,id,args})$ (and, similarly, event type $\funpost$) matches a call to a function 
\textbf{which is not a callback}, having name $\mathit{name}$, \textbf{identifier $\mathit{id}$}, and argument list $\mathit{args}$.
Furthermore, we introduce the new event types $\cbpre$ and $\cbpost$ corresponding to a \emph{call to} and a \emph{return from}, respectively, a callback:
$\cbpre(\mathit{name,id,args})$ (and, similarly, event type $\cbpost$) matches a call to a \textbf{callback}, with name $\mathit{name}$, identifier $\mathit{id}$, and argument list $\mathit{args}$. Matching a callback with its associated asynchronous function call is made possible
by the fact that both calls carry the same identifier.

With this more expressive event domain we can now provide a trace expression $AT$ identifying the correct pattern 
for using functions \lstinline[mathescape=true,basicstyle=\ttfamily\normalsize]{open},  
\lstinline[mathescape=true,basicstyle=\ttfamily\normalsize]{write}, and  \lstinline[mathescape=true,basicstyle=\ttfamily\normalsize]{close} of module \lstinline[mathescape=true,basicstyle=\ttfamily\normalsize]{fs}. Not only the sequence \lstinline[mathescape=true,basicstyle=\ttfamily\normalsize]{open},  
\lstinline[mathescape=true,basicstyle=\ttfamily\normalsize]{write}, and  \lstinline[mathescape=true,basicstyle=\ttfamily\normalsize]{close} must be respected, but also asynchronous operations on the same 
file descriptor must be invoked after the callback associated with the previous operation has been called:
\begin{gather*}
AT=\emptyseq\orop\var{\id}{\opent(\id)\prefixop(CB \shuffleop AT)}\qquad
CB = \var{\fd}{\cb(\id,\fd)\prefixop AW}\\
AW=\var{\id_2}{\writet(\id_2,\fd)\prefixop\cb(\id_2)\prefixop AW} \orop \var{\id_3}{\closet(\id_3,\fd)\prefixop\cb(\id_3)\prefixop\emptyseq}
\end{gather*}

In comparison to the trace expression $PT$ for the synchronous case, the event types $\opent$, $\writet$, and $\closet$ have been extended
to include the identifiers of the corresponding call. Furthermore, 
function calls (event type $\funpre$) rather than returns (event type $\funpost$) are tracked,
because the asynchronous version of \lstinline[mathescape=true,basicstyle=\ttfamily\normalsize]{open} does not return any file descriptor: that value is
retrieved as an argument passed to the associated callback.  

The event type $\cb$ (defined in terms of $\cbpre$) corresponds to calls to callbacks.
It must depend on the corresponding identifier $\id$, and, optionally,
the arguments passed to the callback. In this example, the $\fd$ argument passed to the \lstinline[mathescape=true,basicstyle=\ttfamily\normalsize]{open} callback is considered,
whereas for the callbacks associated with \lstinline[mathescape=true,basicstyle=\ttfamily\normalsize]{write} and \lstinline[mathescape=true,basicstyle=\ttfamily\normalsize]{close} arguments are ignored.

In $AT$ two different binders $\id$ and $\fd$ are required:  
the former captures the identifier associated with a call to \lstinline[mathescape=true,basicstyle=\ttfamily\normalsize]{open}
to be able to match its subsequent callback, while the latter captures the file descriptor on which the callback of \lstinline[mathescape=true,basicstyle=\ttfamily\normalsize]{open} is called. 

Two more binders $\id_2$ and $\id_3$ are required to match the callbacks associated with calls to \lstinline[mathescape=true,basicstyle=\ttfamily\normalsize]{write} and
\lstinline[mathescape=true,basicstyle=\ttfamily\normalsize]{close}.
For clarity, we use three different identifiers $\id$, $\id_2$ and $\id_3$, but we could have equivalently employed
the same name for all three cases, thanks to the scoping rules of the binder construct.

Similarly to what happens for the $PT$ trace expression in the synchronous case, also here recursion and shuffle allow interleaving 
of several events: different calls to \lstinline[mathescape=true,basicstyle=\ttfamily\normalsize]{open}, each uniquely identified by $\id$, and calls to \lstinline[mathescape=true,basicstyle=\ttfamily\normalsize]{write}/\lstinline[mathescape=true,basicstyle=\ttfamily\normalsize]{close} with different file descriptors. 

Finally, we consider an example of a rewriting reduction in case of error.
Let us suppose that a file is correctly opened and then two subsequent writes are performed without waiting for any callback to be executed.
The reduction below accept the trace up-to the first write operation; after that, the second one would be rejected:
\small
\begin{align*}
\mathit{AT} &\trans{e_1} \var{\fd}{\cb(42,\fd)\prefixop AW} \shuffleop \mathit{AT} 
& \{\id \mapsto 42\} &= \mtch(e_1, \opent(\id)) \\
&\trans{e_2} \var{\id_2}{\writet(\id_2,9)\prefixop\cb(\id_2)\prefixop AW} \orop \var{\id_3}{\closet(\id_3,9)\prefixop\cb(\id_3)\prefixop\emptyseq} & \{\fd \mapsto 9\} &= \mtch(e_2, \cb(42, \fd)) \\
&\trans{e_3} (\cb(43)\prefixop AW) \orop \var{\id_3}{\closet(\id_3,9)\prefixop\cb(\id_3)\prefixop\emptyseq} & \{\id_2 \mapsto 43\} &= \mtch(e_3, \writet(\id_2, 9))
\end{align*}
\normalsize
\section{Implementation}
\label{sec:impl}
\begin{figure}[h]
\centering
\includegraphics[keepaspectratio,width=0.7\textwidth]{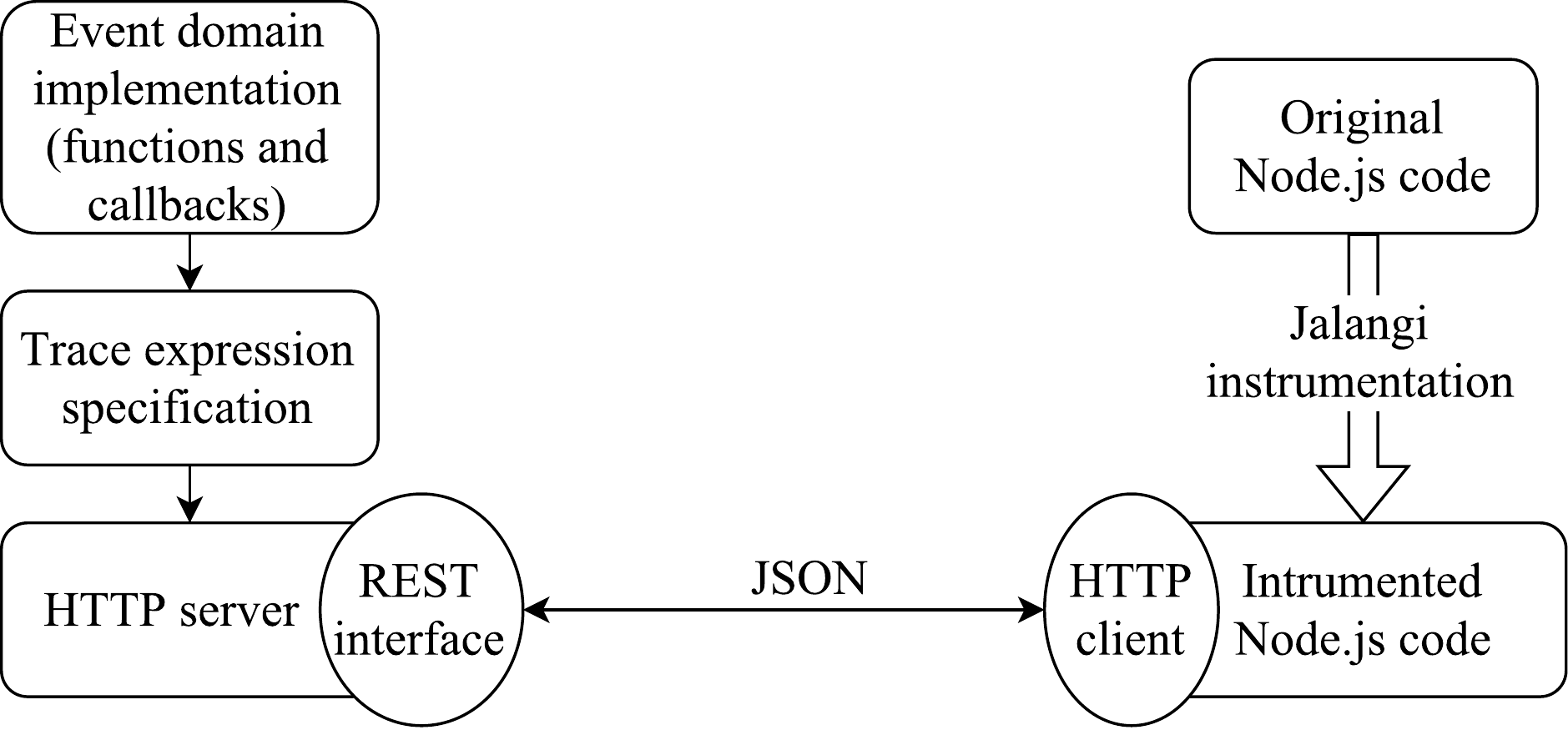}
\caption{High-level view of the implementation.}
\label{fig:overview}
\end{figure}
The implementation of the runtime verification system for Node.js can be divided in two main parts.
On one hand, we (statically) instrument the source code of the program that needs to be verified, adding a piece of code that is able to capture all the relevant events for the domain in use.
On the other hand, we have a Prolog server implementing the operational semantics of trace expressions and offering a simple {REST} interface.
The monitor and the server communicate through {HTTP} requests and responses, effectively implementing a runtime verification system (see Figure \ref{fig:overview}).

We use the standard {JSON} format to exchange data: in our event domain (Node.js functions and callbacks) the instrumented program sends all the information regarding the invoked function, and the monitoring server replies with an error, if any, thus the bidirectional channel.

\subsection{Event domain implementation and source code instrumentation with Jalangi}
Jalangi \cite{Jalangi}\footnote{We refer to Jalangi2 (\url{https://github.com/Samsung/jalangi2}).} is a framework for the implementation of dynamic analyses in JavaScript.
The tool is based on code instrumentation: given a JavaScript program, operations are preceded and followed by calls to Jalangi callbacks, which can be overwritten to implement different analyses.
Instrumented operations include (but are not limited to) accesses to object properties, read and write instructions on variables, loop and conditional statements, and most importantly for us, function and method invocations.

Jalangi offers four callbacks devoted to function invocation analyses:
\begin{itemize}
	\item{\inlinejs{invokeFunPre}} is executed on the call site \emph{before} the function call;
	\item{\inlinejs{invokeFun}} is executed on the call site \emph{after} the function call;
	\item{\inlinejs{functionEnter}} is executed at the \emph{beginning} of the function body;
	\item{\inlinejs{functionExit}} is executed when a \emph{return} statement is evaluated.
\end{itemize}

Let us consider, for instance, the simple Jalangi analysis logging function invocations for Node.js shown in Listing \ref{lst:funclog}.

\begin{lstlisting}[language=JavaScript, caption={Jalangi callbacks for function invocations logging. For the sake of brevity, here we ignore other metadata available to the callbacks as additional arguments.}, label=lst:funclog, float, belowskip=-0.8\baselineskip]
invokeFunPre = function (iid, f, base, args) {
	console.log('going to invoke %s...', f.name)
	return {f: f, base: base, args: args, skip: false}
}
invokeFun = function (iid, f, base, args, result) {
	console.log('after %s...', f.name)
	return {result: result}
}
functionEnter = function (iid, f, dis, args) {
	console.log('beginning of %s...', f.name)
}
functionExit = function (iid, returnVal, wrappedExceptionVal) {
	console.log('returning...')
	return {returnVal: returnVal, wrappedExceptionVal: wrappedExceptionVal, isBacktrack: false}
}
\end{lstlisting}

In the analysis \inlinejs{iid} is the instruction identifier statically associated with any call in the source code,
\inlinejs{f} is the function object that has to be invoked,
\inlinejs{base} and \inlinejs{dis} are the receiver object,
and the other arguments have the obvious meaning.

It is worth noting that most of Jalangi callbacks return the received data to the instrumented code, even if the example above does not exploit this feature.
This allows more complex analysis to actually \emph{change} the behavior of the source code by modifying these values.

Let us now consider a simple Node.js program writing on a file:
\begin{lstlisting}[language=JavaScript,belowskip=-0.8 \baselineskip]
const fs=require('fs')
function foo() {
	fs.writeFile('file.txt', 'Hello, world!', function callback(err) {
		if (err) return console.error(err)
		console.log('The file was saved!')	})
}
foo()
\end{lstlisting}

The output of the Jalangi analysis (Listing \ref{lst:funclog}) when applied to this program would be the following\footnote{We made a few simplifying assumptions here: only logs related to functions \inlinejs{foo} and \inlinejs{fs.writeFile} are shown, and for the moment we assume every function has a name, which is often not the case in JavaScript.}:
{\small
\begin{verbatim}
going to invoke foo...
beginning of foo...
going to invoke writeFile...
after writeFile...
returning...
after foo...
beginning of callback...
returning...
\end{verbatim}
}

The log shows that not all function calls are treated in the same way.
The output makes sense if we recall that Jalangi instruments the given program, not the library code.
As a result, we have three possible scenarios.
For functions both defined in, and called from the program, all the four Jalangi callbacks are executed.
Calls to library functions only get the ``outer'' instrumentation, but we have no information about what is happening inside the function itself.
Viceversa, when an asynchronous library function is completed and a callback defined in the program is executed, only the ``inner'' instrumentation is available.

This approach has advantages, especially regarding Node.js: not only it gives a flexible way to implement a precise analysis, but it also gives the opportunity to track both (asynchronous) library functions and the registered callbacks.
Unfortunately, this also leads to an inconsistency between different kind of functions, as shown in the example above, and some synchronization mechanism is needed between the different instrumentation levels to avoid inconsistencies in the analysis output.

The single-thread nature of Node.js makes this task easier since we can safely assume that only one (instrumented) function at a time will be executed.
Thanks to this we can simply use a stack data structure to record which Jalangi callback has been invoked for each function call in the original program, matching \texttt{invokeFunPre}/\texttt{invokeFun} with \texttt{functionEnter}/\texttt{functionExit} when needed.
This allowed us to build a more abstract and coherent layer over Jalangi, generating the events \(\funpre\) and \(\funpost\) (see previous section) \emph{exactly once} for \emph{every} function in the original program (before and after it, respectively).

\paragraph{(Anonymous) Callbacks}
In the last example some simplifying assumptions were made: all functions were named, and asynchronous callbacks were treated in the same way as other functions.
This is both unrealistic and not so useful for analyzing Node.js applications.
Not only callbacks are usually anonymous functions, but we also want to know exactly when the callback of a specific asynchronous function call gets executed.
Consider the following example:
\begin{lstlisting}[language=JavaScript,belowskip=-0.8 \baselineskip]
const fs=require('fs')
const cb=function() { /* do something */ }
fs.writeFile('hey.txt', 'Hey there!', cb)
fs.writeFile('wow.txt', 'Cool!', cb)
fs.createWriteStream('enough.txt').write('really', cb)
\end{lstlisting}

The callback above is anonymous and it is also used twice with the same asynchronous operation, and once with a different one.
In order to correctly match the asynchronous function calls with their callbacks, our instrumentation creates a unique identifier for each call and substitutes the callback argument with a wrapper where the identifier is stored.
This way, when the callback is actually executed, we can retrieve the information of the original asynchronous call.

At this point we have all the information we need to implement the event domain we used for runtime verification in Section \ref{sec:node}:
\begin{align*}
\funpre(\mathit{name}, \mathit{id}, \mathit{args}) & & \cbpre(\mathit{name}, \mathit{id}, \mathit{args}) \\
\funpost(\mathit{name}, \mathit{id}, \mathit{args}) & & \cbpost(\mathit{name}, \mathit{id}, \mathit{args})
\end{align*}

\subsection{Implementing the monitoring server with SWI-Prolog}
The use of logic programming, and SWI-Prolog \cite{wielemaker:2011:tplp} in particular, for the implementation of the monitoring server offers several advantages, as it allows both the trace expression semantics and the specification of program behavior to be encoded in a natural way.

Our monitoring server implementation is modular: we have developed an HTTP server receiving requests and parsing JSON strings.
Such a server is parameterized with respect to the event domain in use and to the trace expressions encoding the specification that needs to be verified.

\paragraph{How to implement a new trace expression}
We will now show how it is possible to implement a trace expression as a SWI-Prolog module to be loaded by the server, built on top of the function and callbacks event domain.

Since trace expressions are expressed as (recursive) syntactic equations, a crucial feature of SWI-Prolog is its library support for coinduction \cite{SimonMBG06} and regular terms \cite{Courcelle83}.

For instance, let us consider again the parametric trace expression for monitoring asynchronous write operations on multiple files with Node.js:
\begin{gather*}
AT = \emptyseq\orop\var{\id}{\opent(\id)\prefixop(CB \shuffleop AT)} \qquad
CB = \var{\fd}{\cb(\id,\fd)\prefixop AW}\\
AW = \var{\id_2}{\writet(\id_2,\fd)\prefixop\cb(\id_2)\prefixop AW} \orop \var{\id_3}{\closet(\id_3,\fd)\prefixop\cb(\id_3)\prefixop\emptyseq}
\end{gather*}
This can be easily encoded in a SWI-Prolog predicate \texttt{spec} as follows:
\begin{lstlisting}[keywordstyle=\ttfamily,belowskip=-0.8 \baselineskip]
spec(AT) :-
 AT = eps \/ var(ID, open(ID):(CB | AT)),
 CB = var(FD, cb(ID,FD):AW),
 AW = var(ID2, write(ID2,FD):cb(ID2):AW) \/ var(ID3, close(ID3,FD):cb(ID3):eps).
\end{lstlisting}

Note that (parametric) event types can be easily encoded in functional terms (like \texttt{open(FD)}).
In this case, trace expression variables correspond to Prolog logical ones (recall that SWI-Prolog variable names start with an uppercase letter).

The next step is to define a \mtch{} function for the event types in use:
\begin{lstlisting}[keywordstyle=\ttfamily,belowskip=-0.8 \baselineskip]
match(E, open(ID))      :- match(E, func_pre('fs.open', ID, _)).
match(E, write(ID, FD)) :- match(E, func_pre('fs.write', ID, [FD|_])).
match(E, close(ID, FD)) :- match(E, func_pre('fs.close', ID, [FD|_])).
match(E, cb(ID))        :- match(E, cb_pre(_, ID, _)).
match(E, cb(ID, FD))    :- match(E, cb_pre(_, ID, [_,FD|_])).
\end{lstlisting}

New event types can be defined on top of the event domain for functions and callbacks we implemented.
Recall that, for \(\funpre\) and \(\cbpre\), the first argument encodes the name of the function being called, the second one is the unique identifier of the call (and of the registered callbacks, if any), and finally the third one is the list of arguments.
In SWI-Prolog the underscore is used as a placeholder variable (always fresh) for terms we want to ignore.
The syntax \(\mathtt{[}x_1\mathtt{,} \dotsc\mathtt{,} x_n \mathtt{|} l\mathtt{]}\) denotes the list starting with elements \(x_1, \dots, x_n\) followed by the list \(l\).

Finally, the last step consists in exporting the two predicates \texttt{spec} and \texttt{match} so that they can be used by the server:
\begin{lstlisting}[language=prolog,belowskip=-0.8 \baselineskip]
:- module(spec,[spec/1, match/2]).
\end{lstlisting}

With the directive above we created a module named \texttt{spec} exporting the two predicates together with their arity.
The module will then import the event domain implementation in order to get the \mtch{} function implementation for \(\funpre\), \(\funpost\), \(\cbpre\) and \(\cbpost\).

\paragraph{Implementing the event domain}
The event domain implementation works at lower level on the parsed JSON object.
For instance, the following is the implementation of the event type \(\funpre\):
\begin{lstlisting}[language=prolog, keywordstyle=\ttfamily,belowskip=-0.8 \baselineskip]
match(json(O), func_pre(Name, Id, Args)) :-
  member(event='func_pre', O), member(name=Name, O),
  member(id=Id, O), member(args=Args, O).
\end{lstlisting}
Other event domains could be implemented in a similar way.

\paragraph{Operational semantics}
Another advantage of Prolog is that we can almost directly translate the transition rules for the semantics of trace expressions into clauses:

\begin{minipage}{.5\textwidth}
\[
\Rule{and}
{\tau_1\extrans{\ev}\tau_3;\subs_1\quad\tau_2\extrans{\ev}\tau_4;\subs_2}
{\tau_1\andop\tau_2\extrans{\ev}\tau_3\andop\tau_4;\subs}
{\subs=\subs_1\subsMerge\subs_2}
\]
\end{minipage}
\begin{minipage}{.5\textwidth}
\begin{lstlisting}[language=Prolog,keywordstyle=\ttfamily,belowskip=-0.8 \baselineskip]
next(T1/\T2, E, T3/\T4, S) :-
  next(T1, E, T3, S1),
  next(T2, E, T4, S2),
  merge(S1, S2, S).
\end{lstlisting}
\end{minipage}
The merge predicate produces the union of the given substitutions ensuring they are coherent on the intersection of their domains.

\paragraph{HTTP server}
The last piece of the implementation is the HTTP server.
Thanks to the library support of SWI-Prolog and to the conciseness of the language, the HTTP server can be implemented in a few lines of code (for space reasons we do not include the directives importing standard library modules for HTTP and JSON):
\begin{lstlisting}[language=Prolog,keywordstyle=\ttfamily,belowskip=-0.8 \baselineskip]
:- use_module(trace_expressions).
:- use_module(spec).
:- http_handler(/,manage_request,[]).

server(Port) :- http_server(http_dispatch,[port(localhost:Port),workers(1)]).

manage_request(Request) :-
  http_read_json(Request,E),
  nb_getval(state,T1),
  (next(T1,E,T2)->nb_setval(state,T2),reply_json(json([error=(@false)]))
                 ;reply_json(json([error=(@true)]))).
exception(undefined_global_variable,state,retry) :- spec(T),nb_setval(state,T).
\end{lstlisting}

Modules \texttt{trace\_expressions} and \texttt{spec} implement the transition system and the specification, respectively, while \texttt{http\_handler} tells the system that \texttt{manage\_request} should be used to handle requests.

Once an HTTP request is received, first the JSON payload is parsed, then the current state of the server is retrieved, i.e., the current trace expression.
If the incoming event is accepted, then the state is updated with the trace expression yielded by performing a transition step, otherwise an error is detected (\texttt{p->p1;p2} is the SWI-Prolog syntax for conditionals).
The server always replies with a JSON object containing a single boolean-valued field \texttt{error}.
The last line of code deals with the initialization of the server, loading the initial trace expression.

Finally, the server can be run by loading the program above and executing the goal clause \texttt{server(80).} (or any other port).








 






\subsection{Monitoring Node-RED}\label{sec:node-red}

In this section we show how the simple REST interface offered by the SWI-Prolog monitor 
described in the previous subsection can be easily used for monitoring Node-RED flows.

For simplicity, we consider a simple  ``ping-pong'' protocol specified by the
following trace expression:
\begin{align*}
PP&=\var{\val_1}{\ping(\val_1,0)\prefixop T}\\
T&=\var{\val_2}{\pong(\val_2,\val_1)\prefixop\var{\val_1}{\ping(\val_1,\val_2)\prefixop T}}.
\end{align*}
The two employed event types $\ping$, and $\pong$, both depend on pairs of integer values, and are based on
a simple event domain where events are sent messages with two attributes: the type of message
(either ping or pong), and its payload (an integer value).
Event type $\ping(i_1,i_2)$ matches all ping messages sent with payload $i_1$ strictly greater than integer $i_2$, and 
$\pong(i_1,i_2)$ has an analogous definition. Hence, the trace expression $PP$ defines a protocol
consisting of an infinite sequence of alternating ping and pong messages where the first one
must be a ping message with payload strictly larger than 0, while the payload of all other messages must be strictly larger than
the payload of the immediately preceding message.

\begin{figure}
\begin{center}
\includegraphics[keepaspectratio,width=1.0\textwidth]{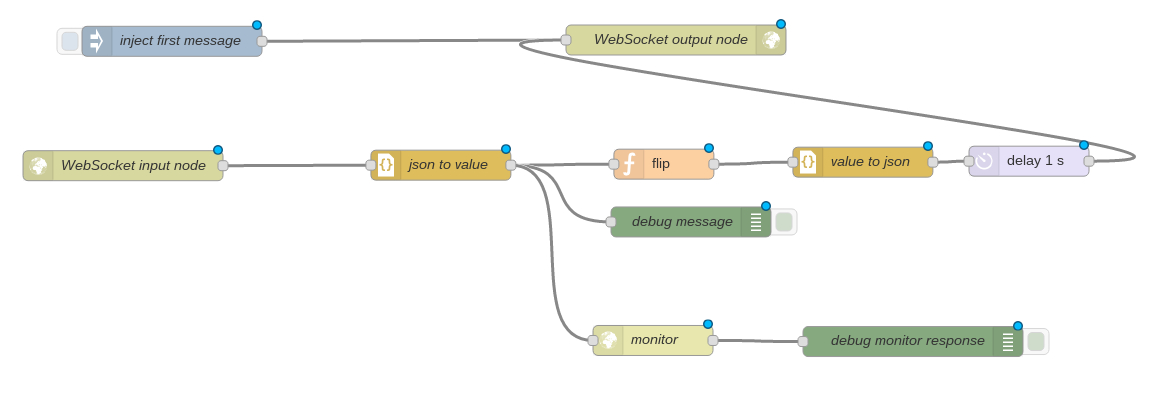}
\end{center}
\caption{A simple Node-RED flow implementing a ``ping-pong'' protocol based on WebSocket}\label{fig:node-red}
\end{figure}

We have implemented such a protocol with the Node-RED flow depicted in Figure~\ref{fig:node-red}, by exploiting the
WebSocket technology. 
All nodes used in the flow are directly offered by the standard library, and have required very simple configuration
settings.

There are two WebSocket nodes, one is an input and the other is an output node, both configured as clients connected\footnote{Available at \url{wss://echo.websocket.org}.} to a simple echo demo server.

The input node receives JSON messages sent from the WebSocket server and send it to the nodes directly connected to it.
In this case, there is a single connected node performing the standard conversion from a JSON string into a corresponding JavaScript value.
The output WebSocket node sends to the echo server the JSON messages arriving at it from the delay node directly connected to it.
The delay node has been inserted simply for limiting the net traffic.

The ``flip'' node is a function node containing simple JavaScript code for transforming the message received at the input, before sending it
to the output.
It flips ping into pong messages, and the other way round, and increases the payload of the message arrived at the input
by a random non negative integer, before sending the transformed message to the output.
The increase of the payload is intentionally allowed to be 0 to simulate errors. 

The message transformed by the ``flip'' node is output to a node which converts it back to the JSON format, before it flows
to the output WebSocket node through the delay node.
Finally, an injection node is responsible for starting the flow with a ping message with payload 1.

Besides the trace expression defined at the beginning of this subsection, monitoring of the flow is simply achieved
by adding\footnote{The flow exhibits also two debugging nodes for logging purposes.} a monitor node, which receives as input the messages that flow throw the WebSocket input node.
Such a node is a standard HTTP request node configured to send to the SWI-Prolog server a post request with the message
received at the input of the node, and to output the corresponding response as a parsed JSON value.


\section{Conclusion and future work}\label{sec:conclu}
This work is a first step towards the use of parametric trace expressions for runtime monitoring of
Node.js applications and Node-RED flows to help detect bugs.
We have developed
a prototype implementation based on Jalangi and SWI-Prolog which offers a simple REST
interface for monitoring IoT systems. The tool supports an event domain
useful for checking the correct use of asynchronous functions and their associated callbacks 
in Node.js applications.
Even if developers may not always have a deep knowledge of the used libraries, which would be required in order to write down a correct specification, ideally the verification system can be deployed together with the library and used by the programmer in a transparent way.

Still a considerable amount of work is needed to assess the approach and prove 
its effectiveness. We are planning to experiment the tool with real Node.js 
applications and to systematically inspect the documentation
of basic modules, such us \lstinline{fs}, \lstinline{http} and \lstinline{express},
on which most Node.js applications rely on, 
to identify patterns of correct use of the exported functions 
that can be expressed with trace expressions, as shown in Section~\ref{sec:node}.
Similarly, we intend to experiment our tool with Node-RED by considering 
actual IoT applications.
 
This study will be essential for identifying scalability issues, and provide
possible solutions.
To this aim, integration of runtime monitoring with
formal verification and testing may prove useful.
Runtime monitoring has been used in conjunction with testing techniques for instance by Artho et al.~\cite{Artho2005209}, 
who propose a framework able to combine automated test case generation, based on systematically exploring the input domain of the program, 
with runtime verification, where execution traces are monitored and verified against properties expressed in temporal logic. 
Deductive software verification~\cite{AhrendtEtAl16} is an interesting technique which seems to offer advantages when employed
together with runtime monitoring.
Regarding static analysis, in the context of parametric verification a dependent type system based on session types has been proposed \cite{Yoshida2010}.

Finally, this work is mainly concerned with logging and error detection, but runtime verification techniques can go further including error recovery procedures,
see for instance ``runtime reflection'' pattern \cite{RuntimeReflection} and the monitor-oriented programming (MOP) methodology \cite{MOP}.
Though Jalangi allows arbitrary code to be executed inside its callbacks, such a possibility would require either a monitor-aware program or some ad-hoc solutions for the system under test.
We leave these considerations for future work.

\bibliographystyle{eptcs}
\bibliography{main}
\end{document}